\documentclass[intlimits,twoside,a4paper]{article}

\usepackage[T2A,T1]{fontenc} 
\usepackage[utf8]{inputenc} 
\usepackage[greek,english]{babel}
\usepackage{teubner}

\usepackage{dsfont}
\usepackage{cmpj3}

\usepackage{graphicx}

\usepackage{color}
\usepackage[normalem]{ulem}

\articletype{A part of the Special collection to the 100th anniversary of the birth of Ihor Yukhnovskii}

\issue{2025}{28}{4}{43501}
\doinumber{10.5488/CMP.28.43501}

\title[50 years of Yukhnovskii's critical point theory]{50 years of Yukhnovskii's critical point theory: its place in the constant flow of theoretical physics}

\author[Yu. Kozitsky]{
			{Yu. Kozitsky}\orcid{0000-0002-4320-8835}
    }
\address{Institute of Mathematics, Maria Curie-Sk\l odowska University, 20-031 Lublin, Poland}


\date{Received September 19, 2025, in final form October 1, 2025}

\begin{document}
\maketitle
\begin{abstract}
Half a century ago, Ihor Yukhnovskii elaborated a method of studying
the critical point of the three-dimensional Ising model based on a layer-by-layer integration in the space of collective variables. His method was an alternative to that based on the $\varepsilon$-expansion for which K.~G. Wilson was awarded the Nobel Prize in Physics in 1982. However, Yukhnovskii's technique, which yielded similar results, provided even deeper insight into the nature of this
phenomenon. At that time, we, professor's students, saw only this aspect of his theory. Later, I realized that the mentioned Yukhnovskii’s work naturally fits into a more general context of the turbulent development of quantum field theory and statistical physics in the last quarter of the twentieth century. The aim of the present article is to look at the main aspects and the impact of Yukhnovskii's theory from this perspective. 
\printkeywords   
\end{abstract}


\section{Introduction}

In 1966, Leo P. Kadanoff in his seminal paper \cite{Kadan} wrote  ``In a recent paper ... Widom has discussed the consequences of the assumption that the free 
energy in a system near a phase transition of second order is a homogeneous function of parameters 
which describe the deviation from the critical point and has shown that this assumption 
leads to consequences which roughly agree with our present numerical information ... about the 
behavior of such systems''. Perhaps, it was this Widom-Kadanoff observation that marked the beginning of the era of scaling and the renormalization group methods in the critical point theory --- first in the Ising model and later in the variety of classical and quantum models, cf. \cite{WF,WK}. Starting from the first half of the 1970s, this theory is being cultivated in Lviv by the group of statistical physicists created and led by professor Ihor Yukhnovskii, who had laid foundations of his original methods based on a layer-by-layer integration in the space of collective variables; see \cite{Yu,YuKP} and the references quoted therein. In this method, the first step made by I. Yukhnovskii and Yu. Rudavskii in \cite{YuRR,YuRRol,YuR} was approximating the Ising spins by some unbounded variables that resemble displacements of ions in anharmonic crystals or lattice quantum fields. 
Also at that time, a group of mathematical physicists made a major breakthrough in constructing interacting quantum fields; see \cite{Simon}. Their idea was to first construct a prototype field as a Markov random field living on the $d+1$ dimensional Euclidean space.
This should be the version of the field of interest living on a $d$ dimensional Euclidean space and in imaginary time. The starting object on this way would be the Gibbs state of the Ising model on a $d+1$ dimensional lattice --- the Ising and the lattice approximation of the mentioned Markov field. Its rigorous construction was already possible in the framework of the Dobrushin-Lanford-Ruelle approach, see \cite{Geor,Simo}, elaborated at the very end of the 1960s.  The next step was taking the limits, eliminating the mentioned approximations, and then passing to the real values of time based on the Osterwalder-Schrader axioms. Significantly, both groups, moving in opposite directions, performed essentially the same steps. This fact,  unknown to the statistical physicist community, 
manifests how high was the level of the research conducted in Lviv at that time. 
Obviously, the mentioned coincidence of steps is far from the only reason for such a conclusion. In fact, Yukhnovskii's layer-by-layer integration approach is one of the independent pillars of modern understanding of the critical point phenomena, which significantly complements other theories of this kind.  
In this article, I aim to bring these thoughts to the attention of the statistical physicist community by looking at the impact and the principal points of Yukhnovskii's theory from today's perspective. 

Fifty years ago the exchange of knowledge was essentially hampered by the Iron Curtain existing between the East and West, as well as by rather modest technical capabilities at that time. Fortunately, these two impediments have now disappeared; however, 
a certain gap still remains in understanding the mathematical aspects of the methods employed. 
This will be taken into account in the present text.        
It should be stressed here that providing any kind of extended review of works in this direction is far beyond my goals.

\section{Basic distribution}

For a system of $n$ physical particles located in a cubic vessel $\Lambda$ in the $d$-dimensional space $\mathds{R}^d$, the corresponding collective variables are defined as Fourier transforms 
\begin{equation}
 \label{1}
 \rho(k) = c_\Lambda \sum_{j=1}^n \exp \left( {\rm i} k\cdot r_j \right), \quad k \in B, \quad {\rm i} = \sqrt{-1 },
\end{equation}
of what would now be called a configuration of particles, fully characterized by their locations $r_1, \dots , r_n \in\Lambda$.  
Here, $B \subset  \mathds{R}^d$ is the domain of the transform, known as the Brillouin zone; $k\cdot r$ stands for the scalar product in $\mathds{R}^d$; $c_\Lambda$ is a suitable normalizing factor.  
The very form of \eqref{1} shows that the collective variables are suitable for studying continuous particle systems such as gases, liquids, etc. In terms of the collective variables, the partition function of the considered system can be presented in the following form
\begin{equation}
 \label{2}
 \Xi_\Lambda = \int \exp \left[ - \beta \Phi_\Lambda(\rho)  \right] J_\Lambda (\rho) \rd^\Lambda \rho,
\end{equation}
where $\Phi_\Lambda(\rho)$ is the configuration energy and $J_\Lambda$ stands for the Jacobian, an essential element of the theory.
The method of collective variables was --- and still is --- a powerful tool in the theory of equilibrium states of interacting particle systems in the continuum, elaborated and applied by I. Yukhnovkii and his followers. Probably, these successes encouraged professor to apply his method to studying the critical point properties of the three dimensional Ising model, which was a very hot topic at that time. However, the first problem he faced on this way was the fact that the corresponding partition function has the form    
\begin{equation}
 \label{3}
 \Xi_\Lambda = \sum_{\dots, \sigma_j = \pm 1 \dots } \prod_{j,j'\in \Lambda}\exp \left[ - \beta \Phi_{jj'}(\sigma_j, \sigma_{j'}) \right],
\end{equation}
i.e., it is rather a finite sum than an integral. The way from \eqref{3} to \eqref{2} was found in \cite{YuRR,YuRRol,YuR}, where it was shown that the substitution 
\begin{equation}
 \label{4}
 \sum_{\sigma_j = \pm 1} g(\sigma_j) \Rightarrow \int_{-\infty}^{+\infty} g(\rho_j) \exp\left[ - P(\rho_j)\right]\rd \rho_j
\end{equation}
does not affect the description of the principal features of the critical point of the three-dimensional Ising model if $P$ is taken in the form
\begin{equation}
 \label{5}
 P(\rho_j) = a_2 \rho_j^2 + a_4 \rho_j^4, \qquad a_4 >0,
\end{equation}
which perfectly corresponds to the interaction term of a quantum field \cite{Simon}. Then, the calculation of the Jacobian $J_\Lambda$ was performed on the bases of \eqref{4} and \eqref{5}, see \cite[equation (13)]{YuR} and/or \cite[equation (5)]{YuKP}.   
Thereafter, the representation of $\Xi_\Lambda$ in the form of \eqref{2} was obtained, which allowed one to apply here a layer-by-layer integration technique, as was mentioned above. In the next section, we turn to its analysis in more detail.  

It should be noticed that the Ising approximation in Euclidean quantum field theory,  i. e., the aforementioned move in the opposite direction as in \eqref{4}, became unnecessary already in 1976 due to the result of \cite{LP}. In that work, the construction of Gibbs states for a lattice model with `unbounded spins' was performed, which opened the possibility to directly employ such states as lattice approximations of the fields under construction. Furthermore, Yukhnovskii's method was later modified in \cite{KYu,Koz} in such a way that the passage in \eqref{4} became unnecessary; see section \ref{S4} below.

\section{Yukhnovskii's method and other models}

\subsection{Posing}

Obviously, most of the methods of the critical point theory of the translation invariant $3D$
Ising model, based on the idea of scaling, are not mathematically strict. Even recognizing the relevant errors, let alone assessing them,  often remains beyond contestation. Nevertheless, such methods are being used during all these years, which might be justified as follows. In the vicinity of the critical point, the only symmetry important for these theories is the asymptotic scale invariance, proper for infinite systems. Then, the mentioned methods artificially introduce --- in one or another way, consciously or not ---  such a symmetry already for finite systems at the cost of losing the mentioned translation invariance and other mathematical details. The way of doing such a `substitution' (also called `ansatz') distinguishes these theories from each other, including their principal outcomes. Similar simplifying ans\"atze  
are welcome not only in the critical point theory. For instance, the naive mean-field theory, cf. \cite[page 216]{Simo}, is widely applied far beyond theoretical physics. 
As R. Kraichnan aptly put it in \cite{Kraich}, by making an ansatz, the initial `true problem' is replaced by  
another one which: (a) reflects the important features of the initial problem; (b) is tractable by methods at hand. Therefore, the quality of a given ansatz-based theory may be assessed by its capability to be reformulated in terms of a new model that meets these criteria. This, in particular, allows one to get a deeper insight into the initial problem as well as to realize the mathematical essence of the ans\"atze on which the theory is based, and thereby to better understand its place among other theories of this kind.

\subsection{Real space renormalization group}

\begin{figure}[!t]
\centering
\includegraphics[width=6cm]{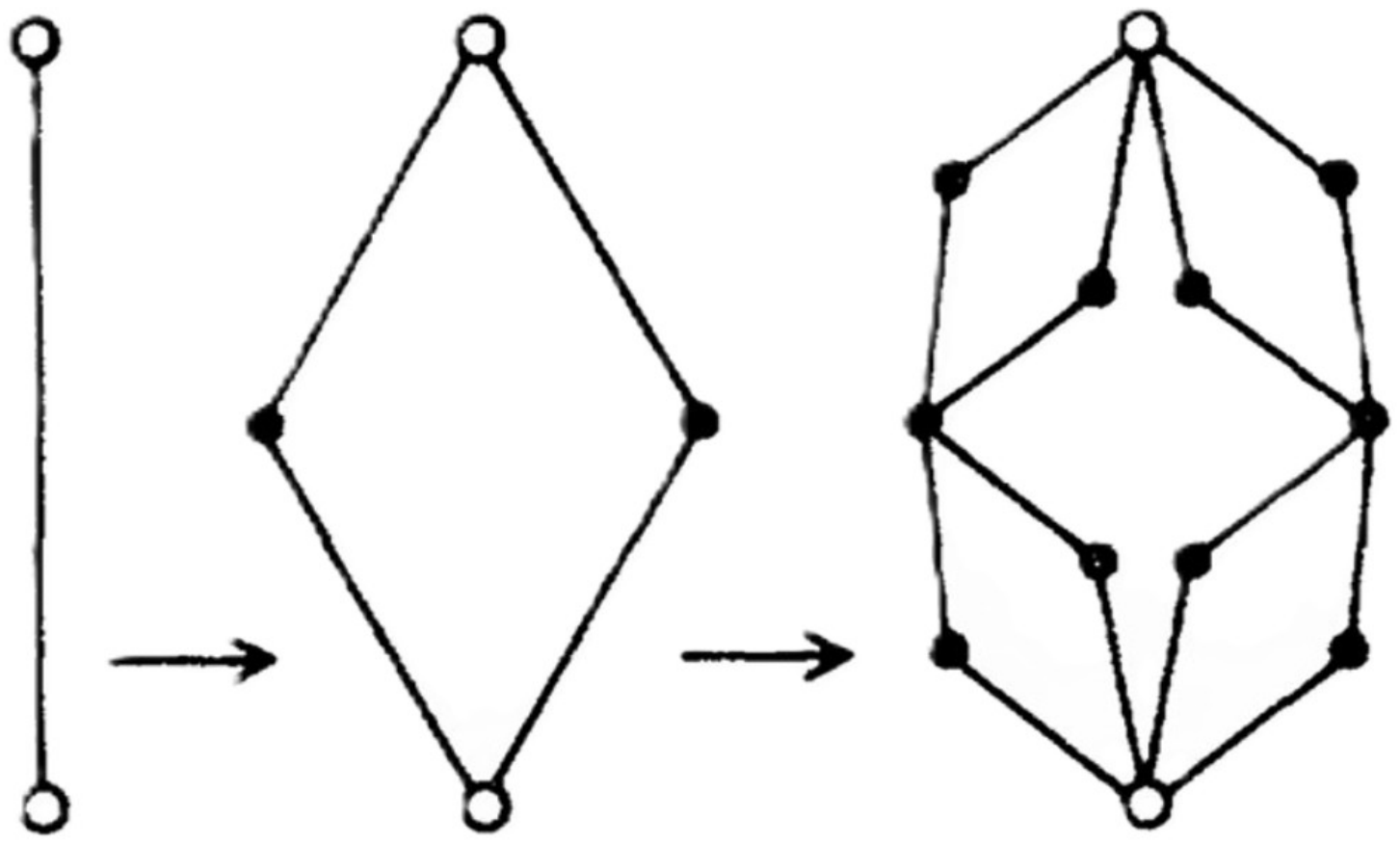}
\caption{Diamond hierarchical lattice, in which bonds are replaced by diamonds.}
\label{fig1}
\end{figure}

\begin{figure}[!t]
\centering
\includegraphics[width=8cm]{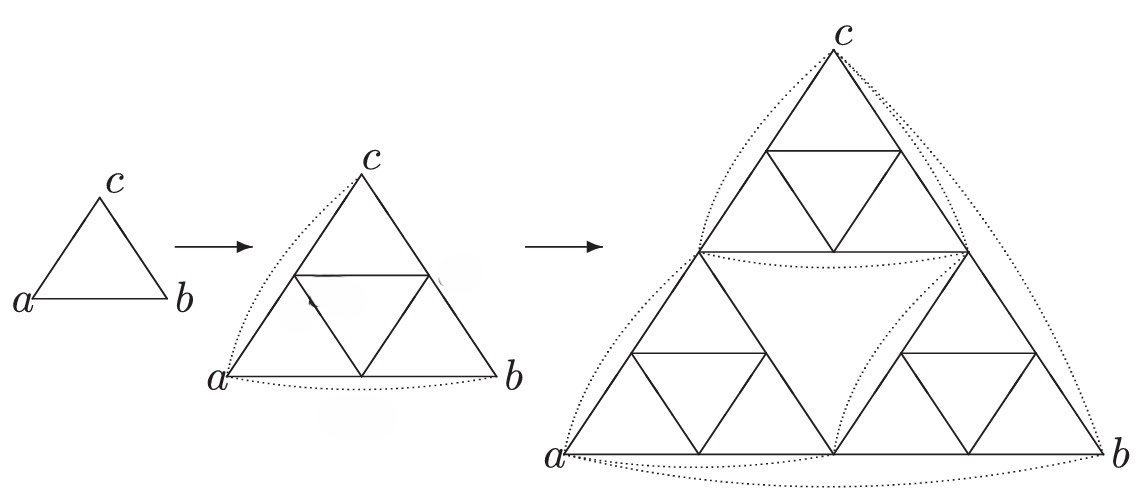}
\caption{Triangle hierarchical lattice, in which nodes are replaced by triangles.}
\label{fig2}
\end{figure}

One of the direct realizations of the aforementioned Widom-Kadanoff
observation was the Migdal-Kadanoff real space renormalization group (RSRG) theory, developed and modified in  \cite{AB,BOP,BO,BZ1,BZ,GKauf,BH,KotK} and in many other publications. Its significant mode of operation is collecting spins into blocks and then collecting blocks into blocks of blocks, etc. However, passing from the distribution of an individual spin to that of a block of spins requires integrating out some degrees of freedom, which for the usual translation invariant cubic lattice is rigorously intractable, and thus can be made by means of ans\"atze. At the same time, the system of such blocks, then blocks of blocks, and so on, manifests a natural hierarchical structure. Thus, the aforementioned simplifying `substitution' in this approach has been performed by replacing the initial translation invariant lattice $\mathds{Z}^d$ by a hierarchical lattice possessing by construction the self-similarity in question. This was first done in
 \cite{BOP,BO}. For the Ising spin model based on such a hierarchical lattice, built in an algorithmic way, the recursion relations of the RSRG theory become exact and tractable. In this case, the Kraichnan-type substitution consists in passing from the Ising model living on a translation invariant lattice to the Ising model living on a hierarchical lattice, built by means of the successive replacement of bonds by diamonds \cite{BOP,BO,BZ,GKauf,BH}, see figure~\ref{fig1} or by the successive replacement of nodes by triangles or other motifs \cite{KoK,KotK}, see figure~\ref{fig2}. Note that motif-based hierarchical lattices have found applications far beyond statistical physics, see, e. g. \cite{LL}.            


 \subsection{Layer-by-layer integration}
 
Now we turn to Yukhnovskii's layer-by-layer integration method restricting ourselves to a very schematic presentation of its main aspects, based mostly on \cite{Koz}. Some mathematical details of this method and its generalizations are outlined in section \ref{S4} below.  

After the ansatz  \eqref{4}, the partition function given in \eqref{3} takes the form
\begin{equation}
 \label{6}
 \Xi_\Lambda = \int_{\mathds{R}^n} \exp\left[\frac{1}{2}\sum_{k\in B} \alpha (k) \rho (k) \rho(-k) - \sum_{l\in \Lambda} P(\rho_l) \right]\prod_{l\in \Lambda} \rd \rho_l,  
\end{equation}
where now $\Lambda$ is a cubic subset of the $d$ dimensional lattice $\mathds{Z}^d$, $B$ is the corresponding Brillouin zone, $P$ is as in \eqref{5}, 
\begin{gather}
 \label{7}
\rho(k) = \frac{1}{\sqrt{n}} \sum_{l\in \Lambda} \rho_l \re^{{\rm i} k\cdot l}, \qquad \alpha (k) = - \beta \sum_{l\in \Lambda} \Phi^\Lambda_{ll'} \re^{-{\rm i} k\cdot (l-l')},  
\end{gather}
and $\Phi^\Lambda_{ll'}$ is obtained from $\Phi_{ll'}$ by imposing periodic conditions on the boundary of $\Lambda$. In view of this, the Brillouin zone $B$ is obtained from the Brillouin zone $(-\piup, \piup]^d$ of the whole lattice $\mathds{Z}^d$ by equidistant partition of each $(-\piup, \piup]$ in such a way that $B$ contains $n$ points. The principal issue here is that both zones are subject to periodic boundary conditions. That is, for $k, k' \in (-\piup, \piup]^d$, $k+k' = (k_1 + k_1', \dots , k_d + k_d')$ where all sums $k_j + k_j'$ are taken modulo $\piup$.  In view of this, $\alpha(k)$ given in \eqref{7} is independent of $l'$ and  
\begin{equation}
 \label{8}
 \Phi^\Lambda_{ll'} = - \beta^{-1} \sum_{k\in B_\Lambda} \alpha(k) \re^{{\rm i} k\cdot (l-l')}.
\end{equation}
We can and will assume that $\alpha(k)$ has a unique maximum at $k=0$, which corresponds to the ferromagnetic nature of the interaction in the model considered.

\begin{figure}[!t]
\centerline{
\includegraphics[width=10.5cm]{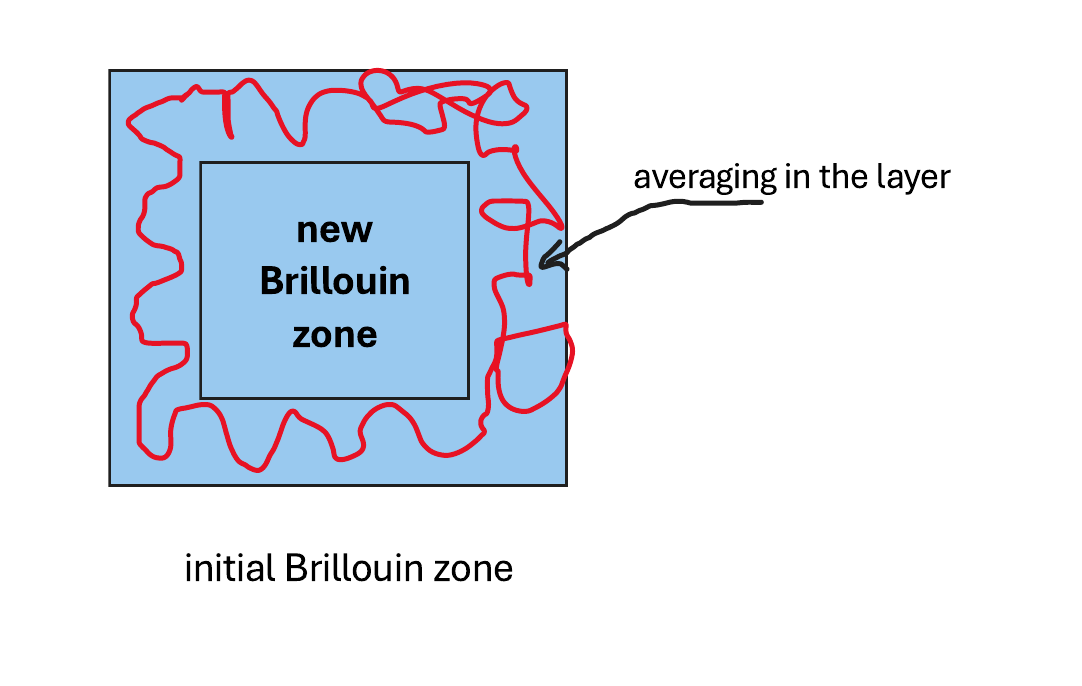}
}
\caption{(Colour online) Averaging in the Yukhnovskii method: (a) the new Brillouin zone $B'$ is cut out from the initial Brillouin zone $B$ with subsequent imposing periodicity on its boundaries; (b) the Fourier image $\alpha(k)$ of the interaction potential is set constant, i. e., $\alpha(k)= \hat{\alpha}_1$ on $B\setminus B'$.}
\label{fig3}
\end{figure}

The key ingredient of Yukhnovskii's method is the following ansatz
\begin{equation}
 \label{9}
 \alpha(k) \Rightarrow [\alpha(k)- \bar{\alpha}_1 ] \mathds{1}_{B'} (k) + \bar{\alpha}_1, \quad k\in B_\Lambda,
\end{equation}
where $B'$ is a rectangular subset of the Brillouin zone $B$, $\mathds{1}_{B'}$ is the indicator of $B'$, i. e., $\mathds{1}_{B'}(k)=1$ for $k\in B'$ and $\mathds{1}_{B'}(k)=0$ otherwise, and $\bar{\alpha}_1$ is the `average' value of the potential $\alpha(k)$ over the layer being the complement of $B'$ in $B$, see figure~\ref{fig3}. More precisely, for a given integer $s>1$,  $B'$ is obtained by imposing the periodicity condition on the boundaries of the set $B \cap (-\piup/s, \piup/s]^d$. Due to such boundary conditions, the addition $k_j + k_j'$ is now set modulo $\piup/s$. It should be stressed here that, for sufficiently large $m$, the summation
\[
k_1 +\cdots + k_m, \quad k_j \in {B}',
\]
calculated in $B$ and in $B'$ may yield different results. 
Now, we set    
\[
 \alpha_1 (k) = \alpha(k)- \bar{\alpha}_1,
\]
employ \eqref{9} in \eqref{6}, and end up with  
\begin{equation}
 \label{10}
 \Xi_\Lambda= \int_{\mathds{R}^n} \exp\left(\frac{1}{2}\sum_{k\in B'} \alpha_1 (k) \rho (k) \rho(-k) - \sum_{l\in \Lambda} [-\frac{\bar{\alpha}_1}{2} \rho_l^2 + P(\rho_l)] \right)\prod_{l\in \Lambda} \rd \rho_l,
\end{equation}
where the following identity was taken into account
\begin{equation}
 \label{11}
 \sum_{k\in B}  \rho(k) \rho(-k) = \sum_{l\in \Lambda}  \rho_l^2,
\end{equation}
which readily follows from the first equality in \eqref{7}. Note that $\Xi_\Lambda$ in \eqref{10} is obtained from that in \eqref{6} by the ansatz \eqref{9} consisting of two ingredients: (a)  averaging the potential; (b) imposing periodicity on the boundaries of $B'$. Hence, the two partition functions are not equal.

Given the periodic conditions imposed on the boundaries of $(-\piup/s, \piup/s]^d$ and $B'$, one can use these sets as Brillouin zones for the new cubic lattice $(s\mathds{Z})^d$ and $\Lambda' \subset (s\mathds{Z})^d$; the latter is subject to periodic boundary conditions. Let $n'$ be the number of nodes contained in $\Lambda'$, which is equal to the number of points contained in $B'$. Then, for $k\in B'$, one can write, cf. \eqref{7},
\begin{equation}
 \label{12}
 \rho(k)  = \frac{1}{\sqrt{n'}} \sum_{l\in \Lambda'} \rho_l \re^{{\rm i}k\cdot l}, \qquad \rho_l = \frac{1}{\sqrt{n'}} \sum_{k\in B'} \rho(k) \re^{-{\rm i}k\cdot l}.
\end{equation}
By \eqref{12} it follows that the first term in $\exp(\cdots)$ in \eqref{10} can be rewritten as a function of $\rho_l$ with $l\in \Lambda'$. Then, it might be natural to rewrite $\Xi_\Lambda$ in terms of the integral over these variables. This was realized in~\cite{Yu,YuKP} with the help of a combination of quite complicated `ans\"atze' with the following result 
\begin{equation}
 \label{10a}
 \Xi_\Lambda= \int_{\mathds{R}^{n'}} \exp\left(\frac{1}{2}\sum_{k\in B'} \alpha_1 (k) \rho (k) \rho(-k) - \sum_{l\in \Lambda'}  P'(\rho_l) \right)\prod_{l\in \Lambda'} \rd \rho_l,
\end{equation}
with , cf. \eqref{5},
\begin{equation}
 \label{10b}
 P'(\rho_l) = a'_2 \rho_l^2 + a'_4 \rho_l^4,
\end{equation}
where $a'_2$, $a'_4$ are certain functions of $a_2$, $a_4$ and $\bar{\alpha}_1$. Again, the partition functions given in \eqref{10a} and \eqref{10} are not equal.  
Then, the described procedure is repeated: one takes $B'' \subset B'$ and makes it the Brillouin zone for $\Lambda'' \subset (s^2 \mathds{Z})^d$. Next, similarly as in \eqref{9}, one employs the ansatz
\begin{equation}
 \label{13}
 \alpha_1 (k) \Rightarrow \alpha_2 (k) \mathds{1}_{B''}(k) + \bar{\alpha}_2 - \bar{\alpha}_1, \qquad k\in B',  
\end{equation}
where $\bar{\alpha}_2$ is the average value of $\alpha(k)$ on the layer $B' \setminus B''$, cf. figure~\ref{fig3}, and 
\[
 \alpha_2 (k) = \alpha (k) - \bar{\alpha}_2.
\]
Now, we make use of the ansatz \eqref{13} in \eqref{10}, similarly as in passing from \eqref{6} to \eqref{10}. 
In its result, the first term in $\exp(\ldots)$ in \eqref{10} is transformed according to
\begin{equation}
 \label{14}
 \frac{1}{2}\sum_{k\in B'} \alpha_1 (k) \rho (k) \rho(-k) \Rightarrow \frac{1}{2}\sum_{k\in B''} \alpha_2 (k) \rho (k) \rho(-k) + \frac{\bar{\alpha}_2 - \bar{\alpha}_1}{2} \sum_{l\in \Lambda'} \rho_l^2, 
\end{equation}
see \eqref{12}. Repeating the same `ans\"atze' that led to \eqref{10a} we arrive at
\begin{equation}
 \label{10c}
 \Xi_\Lambda= \int_{\mathds{R}^{n''}} \exp\left(\frac{1}{2}\sum_{k\in B''} \alpha_2 (k) \rho (k) \rho(-k) - \sum_{l\in \Lambda''}  [a_2'' \rho_l^2 + a_4'' \rho_l^4] \right)\prod_{l\in \Lambda''} \rd \rho_l,
\end{equation}
with $a''_2$, $a''_4$ being certain known functions of $a'_2$, $a'_4$, $\bar{\alpha}_1$ and $\bar{\alpha}_2$. Then, the study of the critical point amounts to studying the asymptotic properties of the map that sends the consecutive pair $(a_2,a_4)$ of the coefficients, cf. \eqref{10b}, to its next iterate, which obviously depends on the asymptotic properties of the averages $\bar{\alpha}_n$. From the viewpoint of the current analysis, it is important to notice that the procedure just outlined creates a hierarchy of translation invariant cubic lattices corresponding to the consecutive Brillouin zones obtained as depicted in figure~\ref{fig3}. In the one dimensional case, this hierarchy is depicted in figure~\ref{fig4}. In particular, the bottom row of black dots presents the initial lattice $\mathds{Z}$ with lattice parameter $c=1$. Its Brillouin zone is $(-\piup, \piup]$, and the set $B$ mentioned above is the subset of this set consisting of $n$ equidistant points -- the Brillouin zone for the set $\Lambda =\{0, 1, \dots , n-1\}\subset \mathds{Z}$ endowed with the periodic boundary condition. The next Brillouin zone $(-\piup/3, \piup/3]$ is obtained by imposing periodicity on the boundaries, and $B'$ is its subset, obtained from $B$ in the same manner.  The former set corresponds to the lattice $3\mathds{Z}$ with lattice parameter $c'=3$. In figure~\ref{fig4}, it is presented by the red lines as lattice sites. Repeating this procedure, one gets $(-\piup/9, \piup/9]$, which is the Brillouin zone for the next lattice $9\mathds{Z}$ with lattice parameter $c''=9$, as well as the corresponding $B''$. The lattice $9\mathds{Z}$ is represented in figure~\ref{fig4} by the top row of black lines as lattice sites. 
Then, the layer-by-layer integration procedure amounts to moving up in this hierarchy of lattices. It is noteworthy that each lattice site at a given hierarchy level can naturally be viewed as the block comprising $s$ lattice sites of the previous level.  
This observation points to the way of formalizing Yukhnovskii's layer-by-layer integration technique. It turns out that the spin model, which can be a candidate for this, was put forward by F. Dyson \cite{Dyson} even before I. Yukhnovskii started working on his method. In the next section, we analyze this connection in more detail.

\section{Dyson's model and Yukhnovskii's method} 
\label{S4}

The layer-by-layer integration method discussed in the previous section was later updated in \cite{KYu,Koz}, where the following improvements were made: (a) the restriction of the initial distribution to the case of~\eqref{5} was lifted and thus it can now be taken in a reasonably arbitrary way, including the original $\sigma = \pm 1$ version; (b) the intermediate `ans\"atze' aimed at preserving the form of $P$ as in \eqref{5} were eliminated, which made the procedure less relying on such `simplifications'. Due to these improvements, 
it has become clear that ans\"atze used in Yukhnovskii's theory yield the replacement of the initial problem by that of studying a version of Dyson's hierarchical model.  
In this section, we briefly describe this updated version, mostly at the level of formal Hamiltonians. 
We refer the reader to \cite{Koz,KYu} for a more complete presentation of this and related issues.  

\subsection{Dyson's hierarchical model}

In 1969, F. Dyson introduced in \cite{Dyson} a special version of the ferromagnetic Ising model living on $\mathds{Z}$, in which the long range interaction between spins $\rho_l$ and $\rho_{l'}$ decays as $(1+ d(l,l'))^{-a} $, where $d(l,l')$ is a special metric on $\mathds{Z}$ determined by a hierarchical structure defined thereon. With the help of this model, he studied the possibility of phase transitions in the Ising model living on $\mathds{Z}$ with the ferromagnetic interaction decaying as $(1+ |l-l'|)^{-a}$, $a>1$, see also \cite{Dyson1}. Later, various versions of Dyson's hierarchical model became independent objects of study; see \cite{BM,BS,Koz,KOZ}. The formal Hamiltonian of Dyson's model introduced in \cite{Dyson} is, cf. equations (3.1)--(3.3) therein,
\begin{eqnarray}
-\beta H & = & \sum_{m=0}^\infty 2^{-1 - a m}\sum_{r\in \mathds{Z}} S^2_{m,r}, \nonumber\\ S_{m,r} & = &  \sum_{l\in \Delta^{(m)}_{r}} \sigma_l, \qquad \Delta^{(m)}_{r} = \{l \in \mathds{Z}: (r-1)2^m +1\leqslant  l  \leqslant r 2^m\},
\label{Oct}
\end{eqnarray}
where $\sigma_l=\pm 1$, $l\in  \mathds{Z}$, are the initial spin variables. The local Hamiltonian of the spins in a given $\Delta^{(m)}_{r}$ is obtained from \eqref{Oct} by restricting the corresponding sums, which yields
\begin{eqnarray}
\label{Oct1}
-\beta H_{m,r}  =  \sum_{n=0}^m 2^{-1- an} \sum_{p: \Delta_P^{(n)} \subset \Delta^{(m)}_r} S^2_{n,p}  =  2^{-1- am} S^2_{m,r} - \sum_{p: \Delta_P^{(m-1)} \subset \Delta^{(m)}_r} \beta H_{m-1,p}.
\end{eqnarray} 
The second line in \eqref{Oct1}  establishes a recursion relation between such local Hamiltonians and thereby between the corresponding partition functions. Similarly to the case of hierarchical lattices mentioned above, Dyson's model possesses the self-similarity symmetry already at the local level.

\subsection{The hierarchy of lattices}

\begin{figure}[!t]
\centering
\includegraphics[width=15cm]{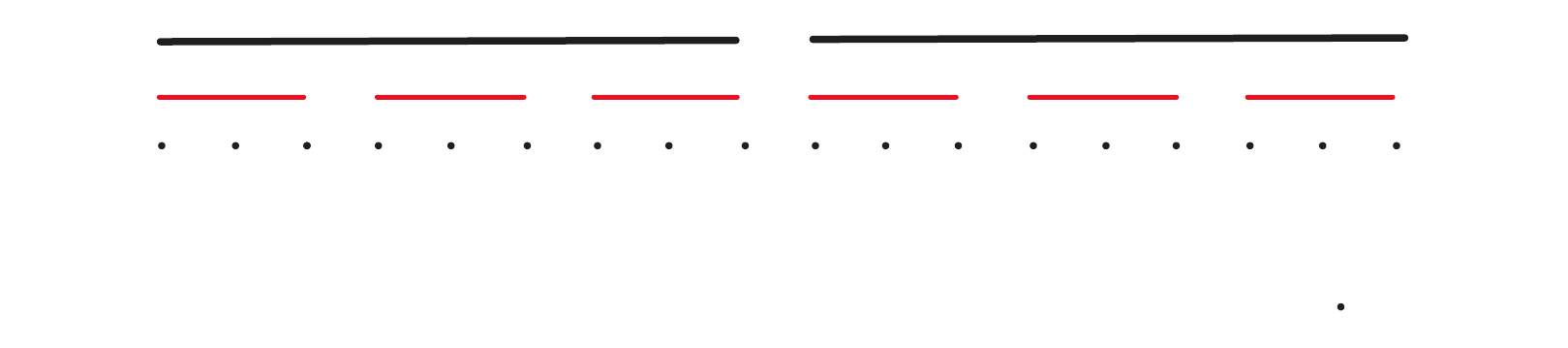}
\caption{(Colour online) Hierarchy of lattices corresponding to the Brillouin zones $B$, $B'$, $B''$.}
\label{fig4}
\end{figure}

As mentioned above, the key idea of Yukhnovskii's method consists in slicing the Brillouin zone $B=(-\piup, \piup]^d$ corresponding to the initial simple cubic lattice $\mathds{Z}^d$ into layers $A_m = B_{m-1}\setminus B_m$, $m\in \mathds{N}$, where   
$B_m = (-\piup/s^m, \piup/s^m]^d$, $s\geqslant 2$ is a fixed integer parameter; cf.  \ref{fig4}, where $s=3$. The aim of this, in particular, is to make each $B_m$ the Brillouin zone of the corresponding cubic lattice by endowing it with the corresponding boundary conditions. Recall that in this section we mostly deal with formal Hamiltonians -- considered as algebraic expressions --- in which summations are taken over the whole lattices.
In view of this, for each $m\in \mathds{N}$, by $B_m$ we denote the Brillouin zone of the lattice $\mathds{Z}^d_m=(s^m\mathds{Z})^d$.

The interaction potential $\Phi_{ll'}$ is assumed to be translation invariant, ferromagnetic and such that
\begin{eqnarray}
\label{F1}
\alpha (k) = -\beta \sum_{l'\in \mathds{Z}^d} \Phi_{ll'} \re^{{\rm i} k\cdot(l-l')}, \qquad k\in B,
\end{eqnarray}
is finite for all $k$. The mentioned condition of ferromagneticity amounts to the following: (a) $\alpha(k)$ attains its maximum at $k=0$ and $\alpha (0) >0$; (b) there exists a sequence of numbers $
\{\hat{\alpha}_m: m \in \mathds{N}\}$ such that 
\begin{equation}
\label{F2}
\inf_{k \in A_m } \alpha (k) \leqslant \hat{\alpha}_m \leqslant \sup_{k \in A_m } \alpha (k), \qquad m \in \mathds{N}.
\end{equation}
Additionally, this sequence should have the following property: there exist $\lambda \in (0,1)$ and $\alpha >0$ such that 
\begin{equation}
  \label{F3}
  \lim_{m\to +\infty} [\alpha (0)- \hat{\alpha}_m]s^{\lambda m d} = \alpha >0.
\end{equation}
Note that a translation invariant interaction potential has these properties whenever it satisfies $\Phi_{ll'}\leqslant 0$ and  
\[
|\Phi_{ll'}| \leqslant \frac{C}{1 + |l-l'|^{d+ \kappa d} }, \qquad \kappa \in (0,1). 
\]
If $\Phi_{ll'}$ decays faster than the upper bound, e. g., if it has finite range, then $\lambda = 2/d$. If it decays 
in the same way, then one takes $\lambda = \kappa$. 

The formal Hamiltonian of the model that we consider can be written in the form
\begin{eqnarray}
 \label{F4}
 -\beta H = - \beta \sum_{l,l' \in \mathds{Z}^d} \Phi_{ll'} \rho_l \rho_{l'} = \int_{(-\piup, \piup]^d} \alpha (k) \rho(k) \rho (-k) \rd k.
\end{eqnarray}
Here, we use the following notations, cf. \eqref{6} and \eqref{8}, 
\begin{eqnarray}
 \rho(k) & = &  \frac{1}{(2\piup)^{d/2}} \sum_{l\in \mathds{Z}^d} \rho_l \re^{{\rm i} k\cdot l},  \nonumber\\
 \rho_l & = &  \frac{1}{(2\piup)^{d/2}} \int_{(-\piup, \piup]^d} \rho (k) \re^{-{\rm i} k\cdot l} \rd k, 
  \label{F5}
\end{eqnarray}
and $\alpha (k)$ is defined in \eqref{F1}. It is noteworthy that the first line in \eqref{F5} is the usual Fourier series, in which $\rho_l$ are the corresponding Fourier coefficients. 

Let the sequence $\{\hat{\alpha}_m: m \in \mathds{N}\}$ be as in 
\eqref{F2} and $\{B_m\}$ be the Brillouin zones as discussed above. Recall that $\mathds{1}_{B_m}$ stands for the corresponding indicator. Thereby, we introduce 
\begin{equation}
 \label{F6}
 \hat{\alpha} (k) = \sum_{m=1}^\infty \hat{\alpha}_m \left[\mathds{1}_{B_{m-1}} (k) - \mathds{1}_{B_{m}} (k) \right]=
 \sum_{m=0}^\infty [\hat{\alpha}_{m+1} - \hat{\alpha}_m ] \mathds{1}_{B_m} (k),
\end{equation}
with $B_0=B$, $\hat{\alpha}_0 =0$. Now, the first ansatz in Yukhnovskii's layer-by-layer integration method consists in replacing the original model  
corresponding to \eqref{F4} by its piecewise constant approximation
 \begin{eqnarray}
 \label{F7}
 - \beta H  =  \int_B \hat{\alpha}(k) \rho(k) \rho(-k) \rd k  
  =   \frac{1}{2} \sum_{m=0}^\infty [\hat{\alpha}_{m+1} - \hat{\alpha}_m ]\int_{B_m} \rho(k) \rho(-k) \rd k,
 \end{eqnarray}
which we obtain by means of \eqref{F6}. At this point, we recall once again that the Hamiltonians \eqref{F7} and~\eqref{F4} are not equal. 
We also recall that each $B_m$ is the Brillouin zone for the lattice $\mathds{Z}^d_m =(s^m\mathds{Z})^d$. Then, we set $\rho^{(m)} (k) = s^{md/2}\rho(k)$, $k\in B_m$, and also, cf. \eqref{F5}, 
\begin{eqnarray}
\label{F8}
\rho^{(m)}_l = \left(\frac{s^m}{2\piup}\right)^{d/2} \int_{B_m} \rho^{(m)}(k) \re^{- {\rm i} k \cdot l} \rd k, \qquad l \in \mathds{Z}_m^d.
\end{eqnarray}
Similarly to \eqref{11} by \eqref{F8} we get
\begin{equation}
 \label{F9}
 \int_{B_m} \rho^{(m)}(k) \rho^{(m)}(-k) \rd k = \sum_{l\in  \mathds{Z}_m^d} [ \rho^{(m)}_l]^2,
\end{equation}
which allows one to rewrite \eqref{F7} in the form
\begin{equation}
 \label{F10}
 -\beta H = \sum_{m=0}^\infty s^{-md}[\hat{\alpha}_{m+1} - \hat{\alpha}_m ] \sum_{l\in  \mathds{Z}_m^d}[ \rho^{(m)}_l]^2.
\end{equation}

\subsection{Block spins}

Now, we have to establish the connections between $\rho^{(m)}_l$ with different $m$. Let $\Lambda^{(m)}\subset \mathds{Z}^d$ be a cubic subset containing $s^m$ lattice sites. We also suppose that, for all $m\geqslant 1$, the shifts 
\begin{equation}
 \label{V}
\Lambda^{(m)}_l = \{ x+ l: x\in \Lambda^{(m)}\}, \qquad l\in \mathds{Z}^d_m, \qquad \Lambda^{(m)}_0 = \Lambda^{(m)}, 
\end{equation}
cover $\mathds{Z}^d$. That is
\begin{equation}
    \label{F10a}
    \mathds{Z}^d = \bigcup_{l\in \mathds{Z}^d_m} \Lambda^{(m)}_l.  
\end{equation}
For $p\leqslant m$ we then set
\begin{equation}
 \label{F11}
 V_{l}^{(m,p)} = \Lambda^{(m)}_l \cap \mathds{Z}^d_p, \qquad l\in \mathds{Z}^d_m.
\end{equation}
By the very definition \eqref{V} it follows that
\begin{equation}
\label{F12} 
 V_{l}^{(m,p)} =  V_{0}^{(m,p)} + l =\{x+l: x\in  V_{0}^{(m,p)} \}.
\end{equation}
Now, by \eqref{F8} we get
\begin{eqnarray}
\label{F13}
\rho^{(m)}_l & = & \left(\frac{s^m}{2\piup}\right)^d \sum_{r\in \mathds{Z}^d_p} \rho^{(p)}_r \int_{B_m} \re^{{\rm i}k\cdot (x-l)} \rd k  \nonumber\\ 
& = &\left(\frac{s^m}{2\piup}\right)^d \sum_{l'\in \mathds{Z}^d_m} \int_{B_m} \bigg(\sum_{x\in V_{0}^{(m,p)}} \rho_{x+l'}^{(p)} \re^{{\rm i}k\cdot x} \bigg) \re^{{\rm i} k\cdot (l'-l)} \rd k
 \nonumber\\ & = & \left(\frac{s^m}{2\piup}\right)^d \sum_{l'\in \mathds{Z}^d_m} \int_{B_m}\eta^{(m,p)}_{l'} (k) \re^{{\rm i} k\cdot (l'-l)} \rd k.
\end{eqnarray}
Now, we employ one more ansatz 
\begin{equation}
 \label{F14}
 \eta^{(m,p)}_{l'} (k) \Rightarrow \eta^{(m,p)}_{l'} (0), 
\end{equation}
which yields in \eqref{F13}
\begin{equation}
    \label{F15}
 \rho^{(m)}_l  \Rightarrow    \eta^{(m,p)}_{l'} (0) = \sum_{x\in V_{0}^{(m,p)}} \rho_{x+l}^{(p)} = \sum_{x\in V_{l}^{(m,p)}} \rho_{x}^{(p)} = \sum_{x\in \Lambda_l^{(m)}} \rho_{x}. 
\end{equation}
The latter means that $\rho^{(m)}_l$ is just the total spin of the block $\Lambda^{(m)}_l$.
By this conclusion the expression in \eqref{F10} becomes the formal Hamiltonian of Dyson's hierarchical model,  introduced in \cite{Dyson} and studied in~\cite{BS,Koz,KOZ}.  

Let now $\Lambda^{(m)}_0$ be as in \eqref{V}. The corresponding local Hamiltonian then is obtained from \eqref{F10}
\begin{equation*}
  -\beta H_{m,0} = \frac{1}{2}\sum_{p=0}^m s^{-pd}[\hat{\alpha}_{p+1} - \hat{\alpha}_p ] \sum_{l\in \Lambda^{(m)}_0\cap  \mathds{Z}_p^d}[ \rho^{(p)}_l]^2.
\end{equation*}
Now, we take into account the ansatz in \eqref{F15} and obtain
\begin{eqnarray}
 \label{F17}   -\beta H_{m,0} = \frac{1}{2}s^{-md}[\hat{\alpha}_{m+1} - \hat{\alpha}_m ] [\rho^{(m)}_0]^2 - \beta \sum_{r\in V_{0}^{(m,m-1)}} H_{m-1,r},
\end{eqnarray}
which is the recurrence relation typical of Dyson's hierarchical model, cf. \eqref{Oct1}. By means of this relation, one can establish the recurrence between the corresponding partition functions.

\subsection{Concluding remarks}
The application of the layer-by-layer integration technique, developed by I. Yukhnovskii and his collaborators in \cite{KYu,Yu,YuKP} and many other works, to studying the critical point of the $d$-dimensional Ising model can be interpreted as the replacement of the initial model by a version of Dyson's hierarchical model performed with the help of the following ans\"atze:
\begin{itemize}
 \item replacement of the Fourier transform of the interaction potential by its piecewise constant approximation  
defined in \eqref{F6};
\item replacement of the cubic subsets $B_m$ of the initial Brillouin zone by the Brillouin zones of the hierarchy of lattices $\{\mathds{Z}^d_m\}$;
\item the ansatz in \eqref{F14}, \eqref{F15}, with the help of which the relationship between the block-spins of different hierarchy levels was established. Its role consists in eliminating the dependence between spins in different blocks other that that arising from the hierarchical structure.  
 
\end{itemize}


\ukrainianpart
\title{50 років теорії Юхновського про критичну точку: її місце у постійній течії теоретичної фізики}

\author{{Ю. Козицький}
}
\address{Інститут математики, Університет Марії Кюрі-Склодовської, 20-031 Люблін, Польща}

\makeukrtitle

	\begin{abstract}
		
		Півстоліття тому Ігор Юхновський розробив метод вивчення критичної точки тривимірної моделі Ізінга, що базується на пошаровому інтегруванні у просторі колективних змінних. Його підхід був альтернативним до методу, що базувався на $\varepsilon$-розкладі, за який К.~Г.~Вільсон отримав Нобелівську премію з фізики у 1982 році. Проте метод Юхновського, який дав подібні результати, забезпечив навіть глибше розуміння природи цього явища. На той час ми, студенти професора, бачили лише цей аспект його теорії. Пізніше я усвідомив, що згадана робота Юхновського природно вписується у більш загальний контекст бурхливого розвитку квантової теорії поля та статистичної фізики останньої чверті ХХ століття. Метою даної статті є розгляд основних аспектів та впливу теорії Юхновського з цієї точки зору.
		
\keywords колективні змінні, ренормгрупа, зона Бріллюена, ієрархічна ґратка, ієрархічна модель Дайсона 
	\end{abstract}

\end{document}